\documentclass[aps,prl,twocolumn,superscriptaddress,preprintnumbers,floatfix,nofootinbib]{revtex4}
\usepackage{graphicx}
\usepackage{hyperref}
\usepackage{epsfig}
\usepackage[usenames,dvipsnames,svgnames,table]{xcolor}

\def\lsim{\raise0.3ex\hbox{$<$\kern-0.75em\raise-1.1ex\hbox{$\sim$}}}
\def\gsim{\raise0.3ex\hbox{$>$\kern-0.75em\raise-1.1ex\hbox{$\sim$}}}

\newcommand{\beq}{\begin{eqnarray}}
\newcommand{\eeq}{\end{eqnarray}}

\newcommand{\bmat}{\left ( \begin{array}{cc}}
\newcommand{\emat}{\end{array} \right )}

\newcommand{\eqref}[1]{(\ref{#1})}

\begin{document}

\title{The Sphaleron Rate in the Minimal Standard Model}

\author{Michela D'Onofrio}
\affiliation{Department of Physics and Helsinki Institute of Physics, PL 64 (Gustaf H\"allstr\"omin katu 2), FI-00014 University of Helsinki, Finland}

\author{Kari Rummukainen}
\affiliation{Department of Physics and Helsinki Institute of Physics, PL 64 (Gustaf H\"allstr\"omin katu 2), FI-00014 University of Helsinki, Finland}

\author{Anders Tranberg}
\affiliation{Faculty of Science and Technology, University of Stavanger, N-4036 Stavanger, Norway}

\date   {\today}

\begin{abstract}
 We use large-scale lattice simulations to compute the rate of baryon number violating processes (the sphaleron rate), the Higgs field expectation value, and the critical temperature in the Standard Model across the electroweak phase transition temperature.  
While there is no true phase transition between the high-temperature symmetric phase and the low-temperature broken phase, the cross-over is sharply defined at $T_c = (159\pm 1)$\,GeV.  The sphaleron rate in the symmetric phase ($T> T_c$) is $\Gamma/T^4 = (18\pm 3)\alpha_W^5$, and in the broken phase in the physically interesting temperature range $130\mbox{\,GeV} < T < T_c$ it can be parametrized as $\log(\Gamma/T^4) = (0.83\pm 0.01)T/{\rm GeV} - (147.7\pm 1.9)$.
The freeze-out temperature in the early Universe, where the Hubble rate 
wins over the baryon number violation rate, is $T_* = (131.7\pm 2.3)$\,GeV.  These values, beyond being intrinsic properties of the Standard Model, are relevant for e.g. low-scale leptogenesis scenarios.
\end {abstract}

\maketitle

\noindent {\it Introduction:} 
The current results from the LHC are in complete agreement with the Standard Model of particle physics:  a Higgs boson with the mass of 125 -- 126\,GeV has been discovered \cite{higgs}, and no evidence of exotic physics has been observed.  
If the Standard Model is indeed the complete description of the physics at the electroweak scale, the electroweak symmetry breaking transition in the early Universe was a smooth cross-over from the symmetric phase at $T > T_c$, where the (expectation value of the) Higgs field was approximately zero, to the broken phase at $T < T_c$\,GeV where it is finite, reaching the experimentally determined value $\langle |\phi| \rangle\simeq 246/\sqrt{2}$ GeV at zero temperature.
The cross-over temperature $T_c$ is somewhat larger than the Higgs mass.
The nature of the transition was settled already in 1995--98 using lattice simulations \cite{KLRS,Gurtler:1997hr,Rummukainen:1998as,Budapest}, which indicate a first-order phase transition at Higgs masses $\lsim\,72$\,GeV, and a cross-over otherwise.

A smooth cross-over means that the standard {\em electroweak baryogenesis} scenarios \cite{EWBG,EWBG2} are ineffective.  These scenarios produce the matter-antimatter asymmetry of the Universe through electroweak physics only, and they require a strong first-order phase transition, with supercooling and associated out-of-equilibrium dynamics.  Thus, the origin of the baryon asymmetry must rely on physics beyond the Standard Model.

Baryogenesis at the electroweak scale is possible in the first place through the existence of the chiral anomaly relating the baryon number of fermions to the topological Chern-Simons number $N_{\rm cs}$ of the electroweak SU(2) gauge fields
\begin{equation}
\Delta N_{\rm cs}(t)
= \frac{1}{32\pi^2}\int_0^tdt' \int d^3x\,
 \epsilon_{\mu\nu\rho\sigma}\textrm{Tr}F^{\mu\nu}F^{\rho\sigma}, 
\label{ncs}
\end{equation}
where $F^{\mu\nu}$ is the SU(2) field strength  \cite{tHooft}. A net change over time of Chern-Simons number leads to a net change in baryon number $B$ (and lepton number $L$), 
\begin{equation}
B(t)-B(0)=L(t)-L(0)=3[N_{\rm cs}(t)-N_{\rm cs}(0)].
\end{equation}
The question is then whether such a permanent change can be achieved through the dynamics around the electroweak transition, either from a symmetric initial state, such as for electroweak baryogenesis, or when it is sourced by another mechanism such as leptogenesis \cite{lepto}, where an initial lepton asymmetry is converted into a baryon asymmetry. 

Close to thermal equilibrium, the evolution of the Chern-Simons number is diffusive, and can be described through the diffusion constant
\begin{equation}
\Gamma = \lim_{V,t\rightarrow \infty} \frac{\langle[ N_{\rm cs}(t)-N_{\rm cs}(0)]^2\rangle}{Vt},
\label{rate}
\end{equation}
also known as the {\it sphaleron rate}. It enters the diffusion equation for lepton and baryon number in baryogenesis \cite{diffusioneq} and leptogenesis (see for instance \cite{burnier}).

The quantity $\Gamma$ has been the focus of extensive work for many years, and a powerful framework and set of analytic and numerical tools have been developed to compute it accurately using non-perturbative lattice simulations (see \cite{us} and references therein). Until very recently, the precise value of the Higgs mass has not been available, although extrapolation of computations at other values of this mass is possible \cite{burnier}. It seems a fitting conclusion to this scientific effort to now employ all the available techniques and finally compute the sphaleron rate of the complete Minimal Standard Model. This will also point forward to similar computations of the rate in extensions of the Standard Model where the electroweak transition can be strongly first order (SM+scalar singlet model, two-Higgs doublet models, Supersymmetric \linebreak models). 

\begin{figure}[t]
  \unitlength1.0cm
  \epsfig{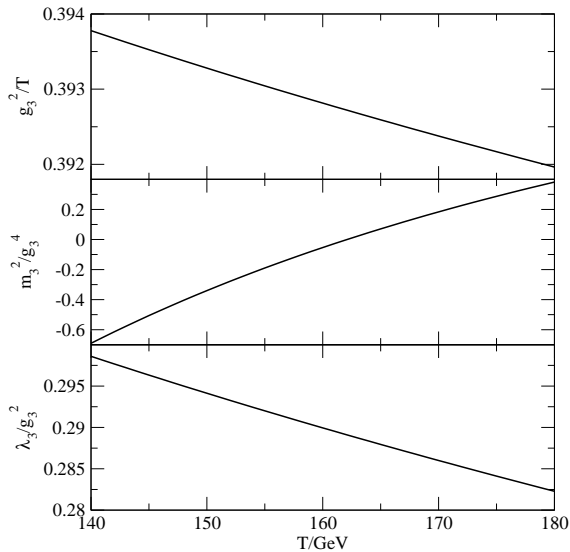}
  \caption{The parameters of the effective theory (\ref{effective}) as
    functions of the temperature.}
  \label{fig:params} 
  \vspace{-0mm}
\end{figure}

{\it Simulation method:}
We will restrict ourselves to a brief summary of the methods and techniques used, and refer the reader to detailed information in the literature (\cite{us} and references therein). 
The full Standard Model is not directly amenable to lattice simulations.
However, at high temperatures the modes corresponding to scales $\ge\,g_{\rm W}T$, including all fermionic modes, can be reliably treated with perturbative methods.  The non-perturbative infrared ($k\,\lsim\,g_{\rm W}^2 T$) physics of the Standard Model is fully contained in
an effective three-dimensional theory, which includes the Higgs field and the spatial 
SU(2) gauge field \cite{generic}:
\begin{equation}
  S=\int d^3x \left[\frac{1}{4}F^a_{ij}F^a_{ij}+
    |D_i\phi|^2 +
    m_3^2|\phi|^2 + \lambda_3 |\phi|^4
\right].
  \label{effective}
\end{equation}
The coefficients $m_3^2$, $g_3^2$ and $\lambda_3$ are functions
of the 4-dimensional continuum parameters ($\alpha_S(M_W)$, $G_F$, $M_{\rm Higgs}$, $M_W$, $M_Z$, $M_{\rm top}$, and the temperature $T$) through a set of 1- and 2-loop matching relations \cite{generic}, and are shown in Fig.~\ref{fig:params} as functions of the temperature.  

We do not include the hypercharge U(1) field explicitly in the effective theory Eq.~\eqref{effective},
because it has little effect on the infrared physics \cite{u1sphaleron,whatsnew}, but we take it into account in our final error analysis. 
Naturally, the U(1) field and the weak mixing angle do contribute to the values of the parameters of Eq.~\eqref{effective}.

The effective action is bosonic and easily discretized on the lattice.
The parameters of the lattice action are perturbatively related to the continuum action \cite{latcont};
we also implement the partial $O(a)$ -improvement of ref.~\cite{Moore:1997np}.
The effective action Eq.~(\ref{effective}) has been very successfully used in calculations
of static thermodynamic properties of the Standard Model, but with unphysical Higgs masses~\cite{KLRS,Gurtler:1997hr,Rummukainen:1998as}.  

For the measurement of the sphaleron rate it is necessary to 
evolve the system in real time.  As such, the effective theory in Eq.~\eqref{effective} does not 
describe dynamical phenomena.  However, the infrared ($k\,\lsim\,g_{\rm W}^2T$) modes have large occupation numbers and behave nearly classically.  Thus, 
it is well motivated to apply classical equations of motion to
Eq.~\eqref{effective} (after introducing canonical momenta).
This method has been used in early studies of the sphaleron rate \cite{early,puregauge}.
However, it has serious problems: due to the UV divergent
Landau damping in the classical theory, the
simulation
results are lattice spacing dependent and the continuum limit does not \linebreak exist~\cite{BMcLS,ASY}. 
These problems can be partially ameliorated by using more complicated effective theories which include so-called hard thermal loop (HTL) effects \cite{htl}, but the continuum 
limit is still out of reach.

A particularly attractive method was first described by B\"odeker \cite{bodeker}: because the dynamics of the infrared modes is fully overdamped, the gauge field evolution can be described with a set of Langevin equations to leading logarithmic accuracy ($\ln^{-1}(1/g_{\rm W})$) \cite{bodeker,ASY}
\begin{equation}
 \partial_t A_i = - \sigma_{\rm el}^{-1} \frac{\partial H}{\partial A_i}
  + \xi_i^a,
\end{equation}
where $\sigma_{\rm el} \approx 0.9239 T$ is the non-Abelian ``color'' conductivity for the Standard Model, and we have identified 
$H/T = S$ in Eq.~\eqref{effective}.  The Gaussian noise vector $\xi$ obeys
\begin{equation}
   \langle \xi_i^a({\bf x},t)\xi_j^b({\bf x}',t')\rangle =
   2\sigma_{\rm el} T \delta_{ij} \delta^{ab} \delta({\bf x}-{\bf x}')
   \delta(t-t').
\end{equation}
The Higgs field is parametrically much less damped than the gauge field, and it can be evolved with Langevin dynamics with a faster rate of evolution \cite{Moore:1998zk}. 
On the lattice, the Langevin dynamics can be substituted with any fully diffusive dynamics,  for instance the heat bath update with random order.  The heat bath update step can be rigorously related to the Langevin time and hence the real evolution time \cite{Moore:1998zk}.  The continuum limit exists and the finite lattice spacing effects have been observed to be modest.

This
method has been successfully used to measure the sphaleron rate in
pure gauge theory \cite{Moore:1998zk} and in the Standard Model
\cite{Moore:2000mx,us}, but not yet using the physical Higgs mass. It
has also been used to study the bubble nucleation rate in first-order electroweak phase
transition~\cite{Moore:2000jw} at unphysically small Higgs mass.

The sphaleron rate is measured using Eqs.~\eqref{ncs}-\eqref{rate}.  However, because topology is not well defined on a coarse lattice, we use the  {\em calibrated cooling} method
of ref.~\cite{Moore:1998swa}, which gives a robust observable for the Chern-Simons number.
In the symmetric phase calibrated cooling can be directly applied to the configurations generated by the heat bath evolution.  Deep in the low-temperature broken phase the situation is more complicated. Although the Langevin dynamics is still correct, the potential barriers between the topological sectors become very large, because the Higgs field has to vanish in the core of the sphaleron.  Hence, the rate becomes very small, and it is not practical to measure it in normal simulations.   This difficulty can be overcome with a special {\em multicanonical} Monte-Carlo computation, where the multicanonical method itself is used to calculate the height of the sphaleron barrier ($\sim$ sphaleron energy), and special real-time runs are performed to calculate the dynamical prefactors of the tunneling process.
The physical rate is then obtained by reweighting the measurements.  For details of this intricate technique, we refer to \cite{us,Moore:1998swa}.
As we will observe, in the temperature range where both methods work, these overlap smoothly.

\noindent {\it Simulation results:}
\begin{figure}[t]
  \unitlength1.0cm
  \epsfig{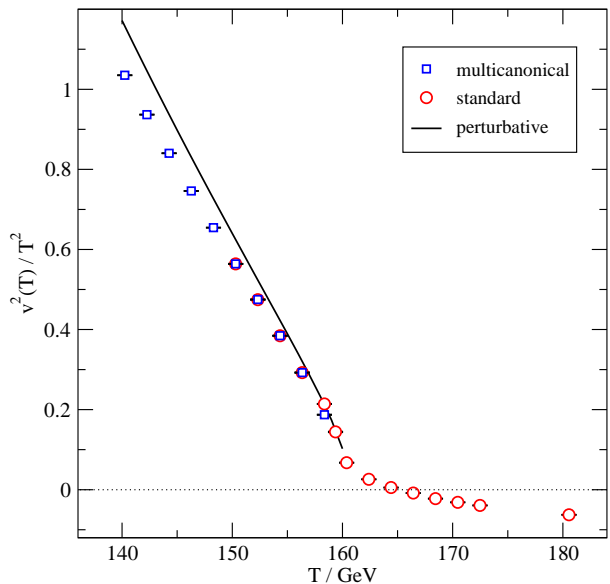}
  \caption{ \label{fig:higgs} The Higgs expectation value as a function of temperature, compared with the perturbative result \cite{KLRS}.}
  \vspace{-0mm}
\end{figure}
We perform the simulations using lattice spacing $a\,=\,4/(9 g_3^2)$ (i.e.~$\beta_G\,=\,4/(g_3^2a)\,=\,9$ in conventional lattice units), and  volume $V\,=\,32^3 a^3$.  In ref.~\cite{us} we observed that the rate measured with this lattice spacing in the symmetric phase
is in practice indistinguishable from the continuum rate, and deep in the broken phase it is within a factor of two of our estimate for the continuum value, well within our accuracy goals.  In fact, algorithmic inefficiencies in multicanonical simulations become severe at significantly smaller lattice spacing, making simulations there very costly in the broken phase.  The simulation volume is large enough for the finite-volume effects to be negligible \cite{us}.

The expectation value of the square of the 
Higgs field, $v^2/T^2 = 2\langle\phi^\dagger\phi\rangle/T$ (here $\phi$ is in 3d units), measures the ``turning on'' of the Higgs mechanism, see Fig.~\ref{fig:higgs}.
As mentioned above, there is no proper phase
transition and $v^2(T)$ behaves smoothly as a function of the temperature.
Nevertheless, the cross-over is rather sharp, and the pseudocritical temperature can be estimated to be
$T_c = 159 \pm 1$\,GeV.
If the temperature is below $T_c$,  $v^2(T)$ is approximately linear in $T$,
and at $T>T_c$, it is close to zero.  The observable 
$\langle \phi^\dagger\phi\rangle$ is 
ultraviolet divergent and is additively renormalized; because of additive renormalization, $v^2(T)$ can become negative.  

We also show
the two-loop RG-improved perturbative result~\cite{KLRS} for $v^2(T)$ in the broken phase. Perturbation theory reproduces $T_c$ perfectly, and $v^2$ is slightly larger than the lattice measurement.  In the continuum limit we expect this difference to decrease
for this observable; in ref.~\cite{us} we extrapolated $v^2(T)$ to the continuum at a few temperature values and with Higgs mass 115\,GeV. The
continuum limit in the broken phase was observed to be about 6\% larger than the result at $\beta_G\,=\,9$.  Thus, for $v^2(T)$ perturbation theory and lattice results match very well.

\begin{figure}[t]
  \unitlength1.0cm
  \epsfig{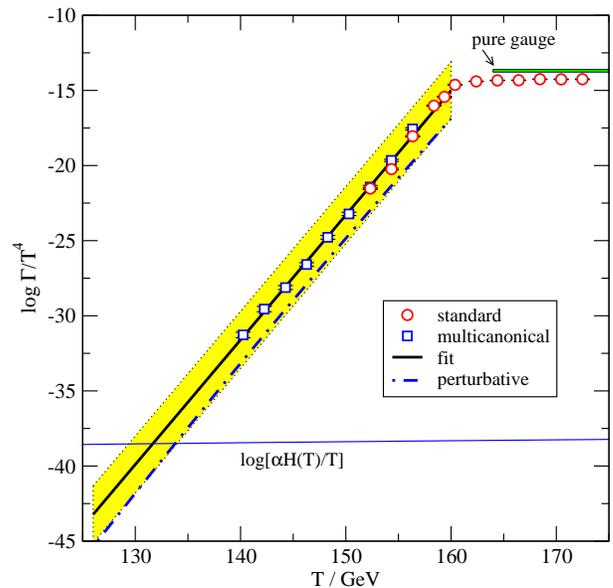}
  \caption{The measured sphaleron rate and the fit to the broken phase rate, Eq.~\eqref{fit}, shown with a shaded error band. The perturbative result is from
Burnier et al.~\cite{burnier} with the non-perturbative correction used there removed; see main text.  Pure gauge refers to the rate in hot SU(2) gauge theory~\cite{puregauge}.
The freeze-out temperature $T_*$ is solved from the crossing of $\Gamma$ and
the appropriately scaled Hubble rate, shown with the almost horizontal line.}
  \label{fig:rate}
  \vspace{-0mm}
\end{figure}

Finally, in Fig.~\ref{fig:rate} we show the sphaleron rate as a function of temperature.  The straightforward Langevin results cover the high-temperature phase, where the rate is not too strongly suppressed by the sphaleron barrier. In fact, we were able to extend the range of the method through the cross-over and into the broken phase, down to relative suppression of $10^{-3}$. 

Using the multicanonical simulation methods we are able to compute the rate 4 orders of magnitude further down into the broken low-temperature phase. The results nicely interpolate with the canonical simulations in the range where both exist.  
In the interval $140\,\lsim\,T\lsim\,155$\,GeV the broken phase rate is very close to a pure exponential, and can be 
parametrized as
\begin{equation}
\log \frac{\Gamma_{\rm Broken}}{T^4} = 
(0.83 \pm 0.01)\frac{T}{{\rm GeV}} - (147.7\pm 1.9).
\label{fit}
\end{equation}
The error in the second constant is completely dominated by
systematics.   We conservatively estimate that
the uncertainties of the leading logarithmic approximation and remaining lattice spacing effects \cite{us} may affect the rate by a factor of two.
The omitted hypercharge U(1) in the effective action (with physical $\theta_W$) 
can change the sphaleron energy by $\approx 1$\% \cite{u1sphaleron} and shift the pseudocritical temperature by $\approx 1$\,GeV \cite{whatsnew}.  These errors have
been added linearly together to obtain the error above.

In the symmetric phase the rate (divided by $T^4$) is approximately constant, and can be
presented as 
\begin{equation}
  \Gamma_{\rm Symm.}/T^4 = (8.0 \pm 1.3) \times 10^{-7} \approx
  (18 \pm 3) \alpha_W^5,
\label{fitsymm}
\end{equation}
where, in the last form, factors of $\ln \alpha_W$ have been absorbed
in the numerical constant.  In pure SU(2) gauge theory
the rate is $\Gamma \approx  (25\,\pm\,2)\,\alpha_W^5 T^4$ \cite{htl,Moore:2000ara}. A difference of this magnitude was also observed in
ref.~\cite{Moore:2000mx}.

In Fig.~\ref{fig:rate} we also show the perturbative result 
calculated by Burnier et al.~\cite{burnier}.  We note that the 
full rate in \cite{burnier} is obtained by including
a large non-perturbative correction to the perturbative rate,
$\log(\Gamma/T^4) = \log(\Gamma_{\rm pert.}/T^4) - (3.6\pm 0.6)$,
where the correction is obtained by matching with earlier
simulations in the broken phase \cite{Moore:1998swa}.  However, these simulations were done with Higgs mass $\approx 50$\,GeV, which is far from the physical one studied here.  With the correction included their result
is a factor of $\approx 150$ below our rate, albeit with large uncertainty.
In Fig.~\ref{fig:rate} we have removed this ad hoc correction altogether,
and the resulting purely perturbative rate agrees with our results well within the given uncertainties of both the lattice and the perturbative computation ($\delta \log\Gamma_{\rm pert.}/T^4 = \pm\,2$). 
 Indeed, by applying a smaller but opposite correction,
$
  \log(\Gamma/T^4) \approx \log(\Gamma_{\rm pert.}/T^4) + 1.6,
$
the central value agrees perfectly with our measurements.
Because the perturbative result is expected to work well deep in the broken phase, the match gives us confidence to extend the range of validity
of our fit \eqref{fit} down to $T\approx 130$\,GeV, in order to cover the physically interesting range.

Finally, we can use the sphaleron rate to estimate when the diffusive sphaleron rate, and hence the baryon number, becomes frozen in the early Universe. The cooling rate of the radiation-dominated Universe is given by the Hubble rate $H(T)$: $\dot T = -H T$.  The freeze-out temperature $T_*$ can now be solved from \cite{burnier}
\begin{equation}
  {\Gamma(T_*)}/{T_*^3} = \alpha H(T_*),
\end{equation}
where $\alpha$ is a function of the Higgs
expectation value $v(T)$, but can be approximated by a constant
$\alpha = 0.1015$
to better than 0.5\% accuracy in the physically relevant
range.  Taking $H^2(T) = \pi^2g^* T^4/(90 M_{\rm Planck}^2)$,
with $g^*=106.75$,\footnote{We neglect $g^*$ changing slightly as the top quark
  becomes massive.}
we find $T_* = (131.7 \pm 2.3)$\,GeV, as shown in Fig.~\ref{fig:rate}.
This temperature enters baryogenesis scenarios where the baryon number is sourced at the electroweak scale, e.g.~low-scale leptogenesis scenarios (see \cite{burnier,Asaka:2005pn} and references therein).   For a more detailed baryon production calculation, the rates
\eqref{fit} and \eqref{fitsymm} can be entered directly into Boltzmann equations.

\noindent {\it Conclusions:} 
The discovery of the Higgs particle of mass 125--126\,GeV enables us to fully determine the properties of the symmetry breaking at high temperatures.  Using lattice simulations of a three-dimensional effective theory, we have located the transition (cross-over) point at $T_c = (159\pm 1)$\,GeV, determined the baryon number violation rate both above and well below the cross-over point, and calculated the baryon freeze-out temperature in the early Universe, $T_* = (131.7\pm 2.3)$\,GeV.  In addition to being intrinsic properties of the Minimal Standard Model,
these results provide input for leptogenesis calculations, in particular for models with electroweak scale leptons. It also provides a benchmark for future computations of the sphaleron rate in extensions of the Standard Model.

\begin{acknowledgments}
  We thank Mikko Laine for discussions.  This work was supported in part by a Villum Young Investigator Grant (AT), by the Magnus Ehrnrooth Foundation (MD) and by the Finnish Academy through grants 1134018 and 1267286. The numerical work was performed using 
the resources at the Finnish IT Center for Science, CSC.
\end{acknowledgments}

\end{document}